\documentclass[final]{aipproc}

\layoutstyle{6x9}

\usepackage{amsmath, amsfonts, amssymb}
\usepackage{graphicx}
\usepackage{soul}
\usepackage{array}
\usepackage[section]{placeins}
\usepackage{wrapfig}

\newcommand{\be}{\begin{equation}}
\newcommand{\ee}{\end{equation}}
\newcommand{\beqn}{\begin{eqnarray}}
\newcommand{\eeqn}{\end{eqnarray}}
\newcommand{\eq}[1]{(\ref{#1})}

\begin{document}

\title[Electromagnetic vacuum superconductivity in magnetic field]{Spontaneous electromagnetic superconductivity 
of vacuum induced by a strong magnetic field: \newline QCD and electroweak theory}

\classification{12.38.-t, 13.40.-f, 74.90.+n}

\keywords{strong magnetic fields, superconductivity, vector meson condensation, QCD}

\author{M. N. Chernodub}{
  address={CNRS, Laboratoire de Math\'ematiques et Physique Th\'eorique, Universit\'e Fran\c{c}ois-Rabelais, F\'ed\'eration Denis Poisson, Parc de Grandmont, 37200 Tours, France},
  altaddress={Department of Physics and Astronomy, University of Gent, Krijgslaan 281, 9000 Gent, Belgium},
}

\author{J. Van Doorsselaere}{
  address={Department of Physics and Astronomy, University of Gent, Krijgslaan 281, 9000 Gent, Belgium}
}

\author{H. Verschelde}{
  address={Department of Physics and Astronomy, University of Gent, Krijgslaan 281, 9000 Gent, Belgium}
}

\begin{abstract}
Both in electroweak theory and QCD, the vacuum in strong magnetic fields develops charged vector condensates once a critical value of the magnetic field is reached. Both 
ground states  have a similar Abrikosov lattice structure and superconducting properties. It is the purpose of these proceedings to put the condensates and their superconducting properties side by side and obtain a global view on this type of condensates. Some peculiar aspects of the superfluidity and backreaction of the condensates are also discussed.
\end{abstract}

\maketitle

\section{Introduction}

Magnetic background fields have gained a lot of interest in high energy physics in the last few years. By carefully studying the vacuum structure of both QCD and electroweak (EW) theory in such backgrounds, it was revealed that 
new very interesting effects may arise. 
In EW theory the pioneering work, which is already 20 years old, has shown that a strong magnetic field leads to a condensation of the $W$ bosons~\cite{Ambjorn:1988tm}. Very recently, it was demonstrated that this $W$-condensed state is in fact, simultaneously, an electromagnetic superconductor and a superfluid~\cite{Chernodub:2012fi}.  At the QCD side, the magnetic field background enhances the chiral symmetry breaking (the magnetic catalysis)~\cite{ref:Igor:review}, affects the chiral and confinement thermal phase transitions~\cite{ref:Marco:review,ref:Eduardo:review} and leads to an electromagnetic superconductivity~\cite{Chernodub:2010qx} and superfluidity of the QCD vacuum~\cite{Chernodub:2011gs}.
The superconducting and superfluid effects in QCD and in the EW theory is the subject of these proceedings\footnote{We would like to stress that we consider pure vacuum only. If matter is present then many other interesting effects may arise: an electric current may be generated along the magnetic field axis in the chirally--imbalanced matter~\cite{Vilenkin:1980fu} (``the chiral magnetic effect''~\cite{Fukushima:2008xe}), the magnetization (electric polarization) of the quark media may depend on electric (magnetic) field background (``the magnetoelectric effect''~\cite{ref:Efrain:review}), the chiral symmetry breaking may be weakened at some density range (``the inverse magnetic catalysis''~\cite{ref:Andreas:review}), and the phase diagram may contain unusual phases exhibiting color superconductivity~\cite{Alford:2001dt}, and phases with certain vector condensates~\cite{ref:vector:dense1,ref:vector:dense2}, including the superfluid/superconducting ones~\cite{ref:vector:dense2}.}.

In a high magnetic field, a nontrivial role in QCD is played by vector mesons.
The corresponding effective action 
contains massive vector fields coupled to the massless photon, resembling strongly EW theory for which a magnetic phase transition was found
in Ref.~\cite{Ambjorn:1988tm}. Both theories turn out to obtain a non-trivial vacuum structure once a critical magnetic background field is reached~\cite{Ambjorn:1988tm,Chernodub:2010qx}. It is possible to calculate the detailed structure of this inhomogeneous vacuum both for QCD and EW theory, with strong similarities between them.

The most surprising effect found in 
the vector-condensed
vacuum state is however its ability to conduct electric current. 
The magnetic field background both
in the QCD vacuum as well as in the EW vacuum, 
leads to a condensation of the corresponding 
charged vector fields and it is exactly the presence of these 
condensates
that allows a current to flow. Needless to say this is at first glance a very counter intuitive claim, 
an electric
current flowing through the vacuum. Moreover, 
this unusual
conductivity  
is dissipationless
so that an analogue of the 
London laws for superconductivity are fulfilled. 

Another interesting effect found both in QCD and in the EW theory in a high magnetic field is the behavior of the neutral vector fields, 
neutral $\rho$ mesons and $W$ bosons, respectively. 
They too do condense, but obtain a more complicated structure 
compared to
their charged counterparts. 
The neutral condensate exhibits 
the same dissipationless motion as the charged condensate in an electric field. This seems to be a very peculiar type of superfluidity.

In the following proceeding we will give a unified view of the vacuum structure and superconducting properties of both the QCD vacuum and the EW vacuum for magnetic backgrounds just above their respective critical values. It must be remarked that these are approximately $10^{16}$ Tesla and $10^{20}$ Tesla, respectively, the rather large discrepancy stemming from the difference between 
the masses of the 
$\rho$ meson and the $W$ boson, correspondingly.

The experimental validation of these calculation is quite hard, but it was 
found theoretically that the magnetic fields higher than the critical value of
$10^{16}$~Tesla (needed for the $\rho$ meson condensation) might be reachable in 
ultraperipheral
heavy-ion collisions at the LHC~\cite{Deng:2012pc}. 
Possible experimental approaches to check the $W$ condensation were also considered in the literature~\cite{Olesen:2012zb}.

\section{Magnetic-field-induced vector condensation}

The $\rho$--meson sector of QCD can be described by the following Lagrangian~\cite{Djukanovic:2005ag}:
\beqn
{\cal L}_{\rho} = -\frac{1}{4} \ F_{\mu\nu}^{2} - \frac{1}{2} \left|D_{[\mu,} \rho_{\nu]}\right|^{2} + m_\rho^2 \left|\rho_\mu\right|^{2}
 - \frac{1}{4} \left| \rho^{(0)}_{\mu\nu}\right|^{2} + \frac{m_\rho^2}{2} \left|\rho_\mu^{(0)} \right|^{2}+\frac{e}{2 g_s} F^{\mu\nu} \rho^{(0)}_{\mu\nu}\,, 
\label{eq:L} 
\label{1}
\eeqn
where the triplet vector field $\rho_{\mu} = (\rho_{\mu}^{(1)},\rho_{\mu}^{(2)},\rho_{\mu}^{(3)} )^{T}$ contains the complex field of the charged $\rho$ meson,
$\rho_\mu = (\rho^{(1)}_\mu - i \rho^{(2)}_\mu)/\sqrt{2}$, and the real-valued field of its neutral counterpart, $\rho^{(0)}_\mu \equiv \rho^{(3)}_\mu$. The $\rho\pi\pi$ coupling $g_s \approx 5.88$ enters the covariant derivative $D_\mu = \partial_\mu + i g_s \rho^{(0)}_\mu - ie A_\mu$ and the strength tensor $\rho^{(0)}_{\mu\nu} = \partial_{[\mu,} \rho^{(0)}_{\nu]} - i g_s \rho^\dagger_{[\mu,} \rho_{\nu]}$.

The Lagrangian of the $\rho$--meson sector of QCD~\eq{eq:L} can be compared to the bosonic sector of the EW Lagrangian,
	\be
{\cal L}_{\mathrm{EW}}=-\frac{1}{4}(W^a_{\mu\nu})^2-\frac{1}{4}X_{\mu\nu}^2+\vert D_\mu\phi\vert^2-V(\vert\phi\vert^2),\qquad D_\mu\phi=\partial_\mu\phi+i\frac{g}{2\cos\theta}Z_\mu\phi,\label{2}
	\ee
where $A_\mu=\cos\theta X_\mu+\sin\theta W^3_\mu$ is the massless photon field, $Z_\mu=-\sin\theta X_\mu+\cos\theta W^3_\mu$ is the massive electrically neutral vector field ($Z$ boson) and $W^-_\mu=(W^1_\mu+iW^2_\mu)/\sqrt{2}$ is the massive charged vector field ($W$ boson). The Higgs field $\phi$ has a nonzero expectation value.

In terms of the Landau level eigenstates (with $n= 0,1,2,\dots$) the quadratic part of Lagrangians \eq{1} and \eq{2} gives us the (squared) energy of the lowest eigenstate around the trivial (with no vector condensates present) minimum:
	\be
	\mathcal{E}^{2} = \langle \psi \vert\hat p_z^2+(2n+1) e B -2e\vec B\cdot\hat s+m^2_{\psi}\vert \psi \rangle
	\equiv p_{z}^{2} + (2 n - 2 s_{z} + 1) e B + m^{2}_{\psi} \,,
	\label{eq:E2}
	\ee
where $\psi = \rho$ (for QCD) and $\psi = W$ (for the EW theory), $m_{\psi}$ and $\vec s$ are the corresponding masses and spins. As the gyromagnetic number is $g = 2$ in both cases, energies lower than the trivial vacuum are possible for the spin projection\footnote{In order to simplify our notation we have assumed that the uniform time-independent magnetic field is directed along the $z$ axis and that $e B > 0$.} $s_{z} = +1$. Moreover, at certain critical magnetic fields, which are different for QCD and EW theory, respectively,
\beqn
B^{\mathrm{QCD}}_{c} = \frac{m_{\rho}^{2}}{e} \approx 10^{16}\, \mathrm{T}\,, \qquad 
B^{\mathrm{EW}}_{c} = \frac{m_{W}^{2}}{e} \approx 1.1 \times 10^{20}\, \mathrm{T}\,, \qquad 
\label{eq:Bc}
\eeqn
the corresponding energies~\eq{eq:E2} become imaginary. This fact signals the presence of an instability for $B>B_{c}$ which allows for a nontrivial vacuum structure to appear both in QCD and in the EW theory.

The charged vector condensate in the ground state has the spin polarization, $s_{z} = +1$. In terms of the fields,  
$\psi_{\mu} = \rho_{\mu}$ (for QCD) and $\psi_{\mu} = W_{\mu}$ (for EW theory) the ground state condensate can be encoded in terms of one a single scalar complex field, $\psi = \rho$ and $\psi = W$, respectively:
	\be
	\psi_{1} = \psi\,, \qquad \psi_{2} = - i \psi\,, \qquad
	\psi_{3} = 0\,, \qquad \psi_{0} = 0\,, 
	\ee
All other polarizations of the charged fields do not condense, so that they can be neglected. The similar statement is true for the neutral fields, their $B$--longitudinal components (i.e. the  timelike component and the component along the magnetic field) are irrelevant in the vacuum, while the $B$--transversal components are nontrivial.

	\begin{figure*}
	\begin{tabular}{cc}\\[5mm]
		\includegraphics[scale=0.22,clip=false]{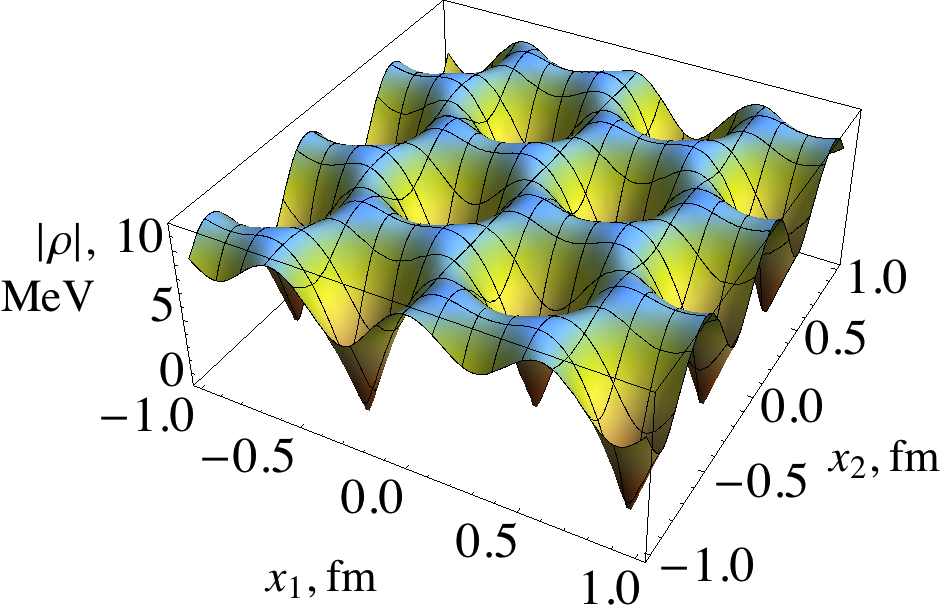} & 
		\includegraphics[scale=0.15,clip=false]{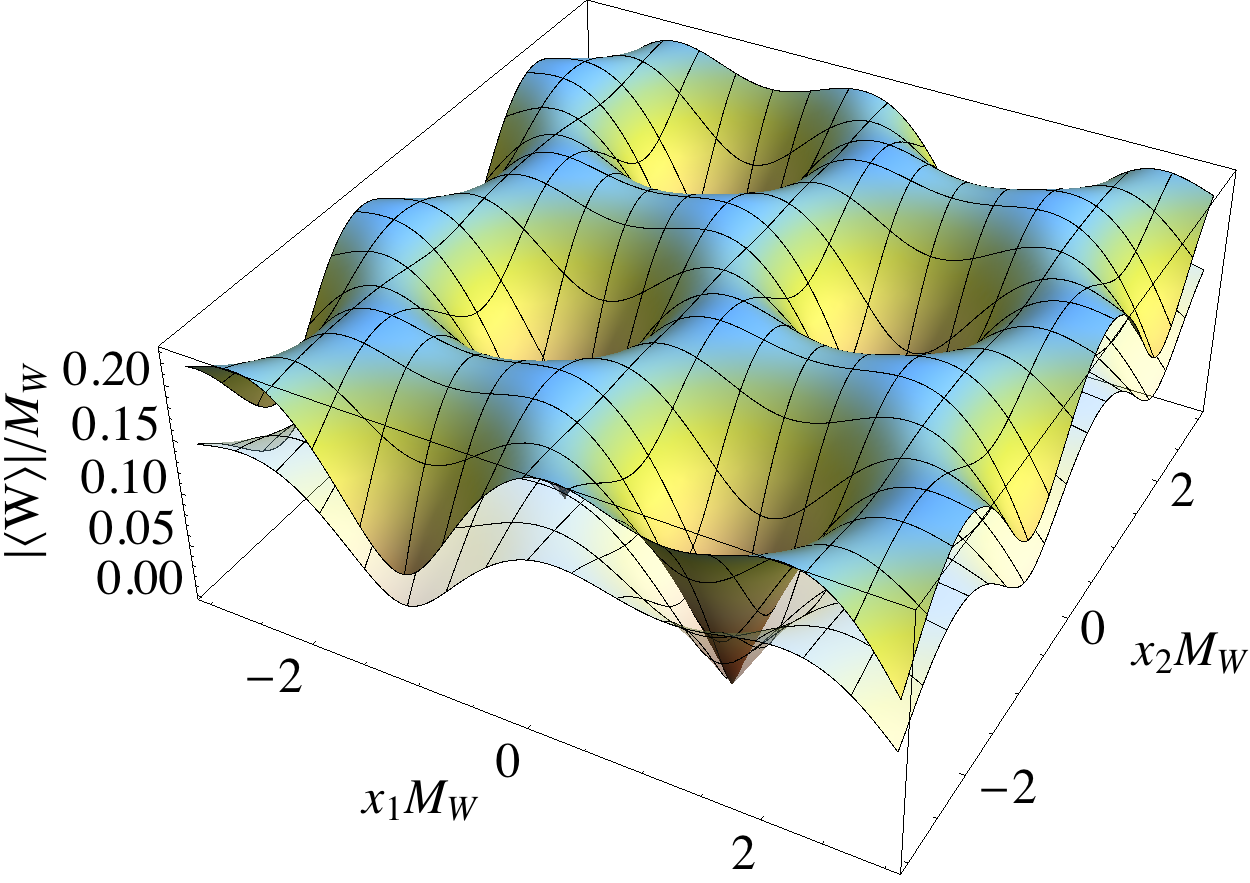}  \\
		\includegraphics[scale=0.30,clip=false]{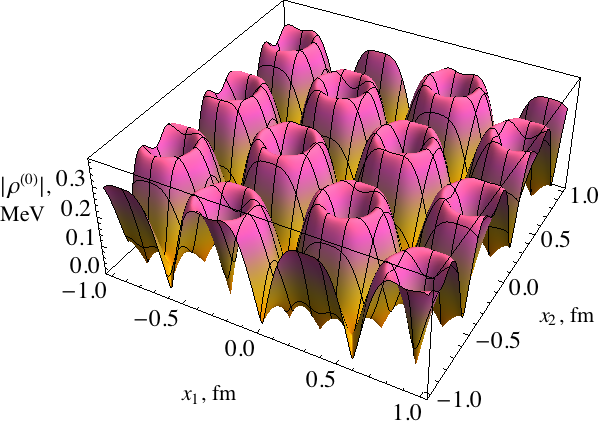} & 
		\includegraphics[scale=0.15,clip=false]{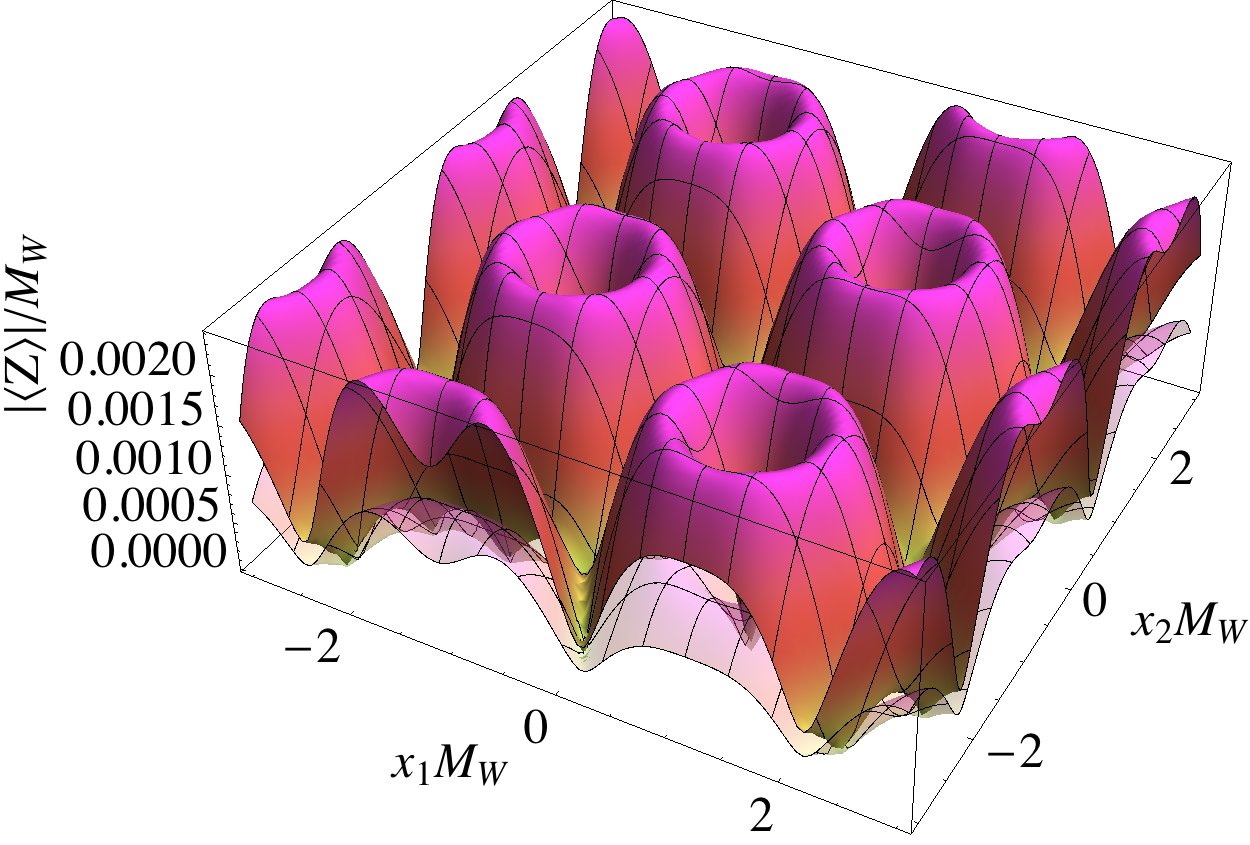} 
	\end{tabular}
		\caption{The absolute values of the charged (the upper panel) and neutral (the lower panel) condensates in the transversal, $(x_{1},x_{2})$, plane. The QCD ($\rho$ and $\rho^{(0)}$) vector condensates are shown at the left panel (taken from Ref.~\cite{Chernodub:2011gs}) while the EW ($W$ and $Z$) vector condensates are shown at the right panel (taken from Ref.~\cite{Chernodub:2012fi}). The solid (semitransparent) surfaces at the right panel correspond to the Higgs mass $M_{H} = 1.1 \, M_{Z}$ ($M_{H} = 2 \, M_{Z}$). The strength of the magnetic field background is just 1\% higher than the corresponding critical magnetic field, Eq.~\eq{eq:Bc}, $B = 1.01 B_{c}$. } 
		\label{fig1}
	\end{figure*}

A minimization of an energy functional corresponding to Lagrangian (\ref{1}) or (\ref{2}) 
gives the 
local 
structure of the relevant vector condensates. It turns out that the condensates in the ground state possess specific topological defects, vortices, which are organized in a hexagonal lattice. The structure of the charged vector fields $\rho$ and $W$ in the ground states of the corresponding theories has the following form~\cite{Chernodub:2010qx,Chernodub:2011gs,Chernodub:2012fi}:
\beqn
    \psi & = & e^{-\frac{i}{2}eBxy}\sum_n C_n e^{-\frac{eB}{2}\left[(x-n\nu L_B)^2-2in\nu yL_B\right]}\,, \qquad \psi = 	W\,, \ \rho\,,
    \label{eq:charged}\\
	 L_B&= & \sqrt{\frac{2\pi}{eB}},\qquad \nu=\sqrt{\frac{\sqrt{3}}{2}},\qquad {C_{2n}\choose C_{2n+1}}={C_0 \choose iC_0}\,,
\eeqn
where the value of the dimensional constant $C_0 = C_{0}(B)$ can be defined numerically via a minimization procedure. The condensates are zero for relatively low magnetic fields, $B < B_{c}$, where the corresponding critical values $B_{c}$ are given in Eq.~\eq{eq:Bc}.

The massive vector fields $\rho^{(0)}$ and $Z$ in the ground state of QCD and EW theory are expressed via the corresponding charged condensates in the ground states~\eq{eq:charged}:
	\beqn
	 \rho^{(0)} & \equiv & \rho^{(0)}_{1} + i \rho^{(0)}_{1} = -2ig_s\frac{\partial_x+i\partial_y}{\Delta_\perp-m_0^2}\vert\rho\vert^2\,, \nonumber \\ 	  
	 Z & \equiv & Z_{1} + i Z_{2} = i\frac{g}{2}\cos\theta_W\frac{\partial_x+i\partial_y}{\Delta_\perp-M_Z^2}\vert W\vert^2\,,
	\eeqn

Thus, the behavior of the charged $(\rho)$ and neutral $(\rho^{(0)})$ vector condensates in the ground state of QCD are very similar to the charged $(W)$ and neutral $(Z)$ condensates in the EW model. This statement is true for relatively small condensates which are realized in the vicinity of the corresponding critical magnetic field $B_c$, Eq.~\eq{eq:Bc}. The difference between these theories appears at stronger fields~\cite{VanDoorsselaere:2012zb} due to the presence of the Higgs field in the EW theory and absence of a fundamental scalar fields in QCD. As the magnetic field increases above $B_{c}^{\mathrm{EW}} \equiv B_{c}^{\mathrm{EW}}$, the Higgs condensate decreases and thereby lowers the mass of all vector fields in the EW model. As a consequence of this effect, the EW vacuum should suffer a second phase transition at another, second critical value of magnetic field $B_{c2}^{\mathrm{EW}} > B_{c1}^{\mathrm{EW}}$. Both the $W$ and $Z$ bosons become massless at this second phase transition, and in some sense one may state that the $SU(2)$ gauge symmetry becomes virtually restored at $B > B_{c2}^{\mathrm{EW}}$ . However, this ``restoration'' just means that the $SU(2)$ symmetry is no longer {\it spontaneously} broken by the Higgs mechanism. The gauge symmetry should anyway be broken {\it explicitly} due the presence of the vector condensates of the $W$ and $Z$ bosons which are induced by the strong magnetic field background. The second phase transition is naturally absent in QCD, which does not include a fundamental scalar (Higgs-like) field.

\section{Vacuum superconductivity and superfluidity}

In order to study the conducting properties of the vacuum state we apply a tiny (probe) electric field along the direction of the magnetic field (i.e., along the $z$ axis in our conventions).
In a conventional superconductor the electric current satisfies the London relation,
	\be
	\frac{\partial J^i}{\partial t}=\frac{n_s e^2}{m} E^i\,, 
	\qquad \mbox{or} \qquad
	\partial_{[0}J_{3]}=-2e^2\vert\phi\vert^2E \,,
        \label{london}
	\ee
where $n_s$ is the density of electric charges in the condensed state, $m$ is their mass. The second relation in Eq.~\eq{london} is the relativistic version of the London equation, where $\phi$ is the field of the condensed superconducting carriers (Cooper pairs). As we will see below, similar relations arise both in QCD and in the EW theory due to the presence of the magnetic--field--induced vector condensates.

For small $W$ condensates in the EW model, the equations of motion can be linearized, and subsequently written via the  conserved electric, $J$, and neutral $Z$-boson, $J^{Z}$, currents: 
\beqn
	J_\mu & = & ie\left(W^{+\mu}W_{\mu\nu}^--W^{-\mu}W_{\mu\nu}^++\partial^\mu(W^{+}_{\mu}W^-_{\nu}-W^{-}_{\mu}W^+_{\nu})\right) = \partial^\nu F_{\nu\mu}\,,
   	 \label{eq:EW:1}\\
	 J^Z_\mu & =& \partial^\nu Z_{\nu\mu}\,, \qquad \qquad \cos\theta_W J_\mu-\sin\theta_W J^Z_\mu + M_Z^2 Z_\mu  = 0\,.
	 \label{eq:EW:2}
\eeqn
In the same approximation one gets~\cite{Chernodub:2012fi} an EW analogue of the London equation~\eq{london}:
	\beqn
	\partial_{[0}J_{3]}=-e^2\vert W (B) \vert^2 E\label{W}\,, \qquad B > B^{\mathrm{EW}}_{c}\,.
	\label{eq:London:EW}
	\eeqn

Doing identical calculations for $\rho$ mesons in QCD, one finds a similar set of equations for the electric current, $J_\mu$, and the neutral current, $J_\mu^{(0)}$, respectively:
\beqn
	J_\mu & = & J^{ch}_\mu+J^{(0)}_\mu = - \frac{e}{g_s}\, m_0^2 \, \rho^{(0)}_\mu =\partial^\nu F_{\nu\mu} ,\\
	J^{(0)}_\mu & = & \frac{e}{g_s}\partial^\nu\partial_{[\nu}\,\rho^{(0)}_{\mu]}, \qquad 
	 \label{eq:QCD:2}
	J^{ch}_\mu = -ie\rho^{\dag\mu}\rho_{\mu\nu}+ie\rho^\mu\rho^\dag_{\mu\nu}-ie\partial^\mu(\rho^\dag_{[\mu}\rho_{\nu ]})\,.
\eeqn
The London--like equation in high-magnetic field phase of QCD is as follows~\cite{Chernodub:2010qx}:
	\be
	\partial_{[0}J_{3]}= -4e^2 m_0^2 \frac{1}{- \Delta_{\perp} + m_0^2}  | \rho|^2E\,, \qquad B > B^{\mathrm{QCD}}_{c}\,,
	\label{eq:London:QCD}
	\ee
where $\Delta_{\perp} = \partial_{1}^{2} + \partial_{2}^{2}$ is the two--dimensional Laplacian in the $B$--transverse directions.

Thus, in the $B$--longitudinal direction (i.e. for the electric current along the magnetic field axis) one gets the exact London relation in the EW model~\eq{eq:London:EW}, where the $W$ condensate superconducts as if it were a field of the Cooper pair~\eq{london}. The London relation holds also in QCD, where the superconductivity coefficient is a nonlocal function of the $\rho$ condensate~\eq{eq:London:QCD}. In the $B$--transverse directions both QCD and the EW model are not superconducting~\cite{Chernodub:2010qx,Chernodub:2012fi}.

In both these theories the neutral condensates behave as an exotic superfluid~\cite{Chernodub:2010qx,Chernodub:2012fi}. Indeed, one can show that, surprisingly, the neutral currents $J^{Z}_{\mu}$ of the $Z$ bosons~\eq{eq:EW:2} and the neutral current $J^{(0)}_{\mu}$ of the $\rho$ mesons~\eq{eq:QCD:2}, satisfy London relations as well:
\beqn
\partial_{[0}J_{3]}^Z & = &-e^2\cot\theta_W\frac{\Delta_\perp}{\Delta_\perp-M_Z^2}\vert W\vert^2E\,, \\
\partial_{[0} J^{(0)}_{3]}&=&4e^2\frac{\Delta_\perp}{\Delta_\perp-m_0^2}\vert \rho\vert^2E\,.
\eeqn

Notice that the emergent vector condensates locally modify both the strong magnetic field~\cite{Chernodub:2010qx,Chernodub:2012fi} and the weak (probe) electric field. In order to see the latter, we get the following general relation from Eq.~\eq{eq:EW:1}:
	\be
	\partial_{[\mu}J_{\nu]}=\partial_{[\mu} (\Box A_{\nu]}-\partial_{\nu]}\partial^\lambda A_\lambda)=\Box F_{\mu\nu}\,.
	\ee
Then we combine this relation with the corresponding London laws, Eq.~\eq{eq:London:EW} and Eq.~\eq{eq:London:QCD}, for the static electric field $E_{3} = E_{3}(x_{1},x_{2})$, homogeneous along the magnetic field direction~$x_{3}$. tIn the weak--condensate approximation we get the following relations in QCD and in EW theory, respectively:
	\be
	\Delta_{\perp} E_{z}=4e^2 m_0^2 \frac{1}{- \Delta_{\perp} + m_0^2}  | \rho|^2E_{z},\qquad \Delta_{\perp} E_{z}=e^2 \vert W\vert^2 E_{z}\,.
	\ee 
Thus, a backreaction from the vector condensates should lead to inhomogeneities of both electric and magnetic external fields.
		
\clearpage

\section{Conclusions}

We have briefly overviewed the basic properties of the magnetic-field-induced $\rho$--meson and $W$--boson vector condensates in, respectively, the vacua of QCD and EW theory. The superconducting properties of these condensates are revealed in a weak (probe) electric field background. The superconductivity current is proportional to the squared value of the corresponding charged condensate, in a close similarity with the London laws of conventional superconductivity in condensed matter physics. In the strong magnetic field the neutral condensates of, respectively, the $\rho^{(0)}$ mesons and of the $Z$--boson appear as well. These neutral condensates turn out to be sensitive to the external electric field, giving a new kind of superfluidity, which exists in combination with the superconducting component only. 

The work MNC was supported by Grant No. ANR-10-JCJC-0408 HYPERMAG.

\bibliographystyle{aipproc}   

\end{document}